\documentclass[conference]{IEEEtran}
\IEEEoverridecommandlockouts
\usepackage{cite}
\usepackage{amsmath, amssymb, amsfonts}
\usepackage{algorithm}
\usepackage{algpseudocode} 
\usepackage{graphicx}
\usepackage{textcomp}
\usepackage{xcolor}
\usepackage{subcaption}
\usepackage{mathtools}
\usepackage{flushend}
\usepackage{stfloats}
\usepackage{multirow}
\usepackage{booktabs}
\usepackage{cleveref}

\crefname{figure}{Fig.}{Figs.}
\crefname{table}{Table}{Tables}
\crefname{algorithm}{Algorithm}{Algorithms}
\crefname{section}{Section}{Sections}

\makeatletter
\let\MYcaption\@makecaption
\makeatother
\IEEEoverridecommandlockouts\IEEEpubid{\makebox[\columnwidth]{978-1-6654-3540-6/22~\copyright~2022 IEEE \hfill} \hspace{\columnsep}\makebox[\columnwidth]{ }}

\begin{document}

\title{Adapting CSI-Guided Imaging Across Diverse Environments: An Experimental Study Leveraging Continuous Learning\\
\thanks{

}
}

\author{\IEEEauthorblockN{
Cheng Chen\IEEEauthorrefmark{1}\IEEEauthorrefmark{2}, 
Shoki Ohta\IEEEauthorrefmark{1}, 
Takayuki Nishio\IEEEauthorrefmark{1}, 
Mohamed Wahib\IEEEauthorrefmark{2},\\
}

\IEEEauthorblockA{
\IEEEauthorrefmark{1}\textit{Tokyo Institute of Technology}, Tokyo, Japan\\
chen.c.aj@m.titech.ac.jp, nishio@ict.e.titech.ac.jp}
\IEEEauthorblockA{
\IEEEauthorrefmark{2}\textit{RIKEN Center for Computational Science}, Kobe, Japan
}

}
\maketitle

\begin{abstract}
This study explores the feasibility of adapting CSI-guided imaging across varied environments. Focusing on continuous model learning through continuous updates, we investigate CSI-Imager's adaptability in dynamically changing settings, specifically transitioning from an office to an industrial environment. Unlike traditional approaches that may require retraining for new environments, our experimental study aims to validate the potential of CSI-guided imaging to maintain accurate imaging performance through Continuous Learning (CL). By conducting experiments across different scenarios and settings, this work contributes to understanding the limitations and capabilities of existing CSI-guided imaging systems in adapting to new environmental contexts.
\end{abstract}


\section{Introduction}

Channel State Information (CSI) represents a pivotal element in wireless communication, elucidating the signal's interaction with its environment by detailing the propagation characteristics across multiple frequencies and subcarriers. In the realm of Orthogonal Frequency-Division Multiplexing (OFDM) systems, this translates to a granular view of each subcarrier's amplitude and phase, reflecting signal strength and propagation delays. The nuanced understanding of frequency-dependent signal behavior afforded by CSI is instrumental for sensing applications, where variations in signal reflection and attenuation can be harnessed to infer environmental contours and objects.

Advancements in wireless sensing technologies have significantly expanded the horizons of imaging capabilities, with CSI-guided imaging standing out as a notable innovation \cite{9919801}, \cite{9380376}, \cite{Chen2022Cross-Domain}, \cite{chen2023trans}. This technology, which utilizes CSI derived from Wi-Fi signals, facilitates imaging in scenarios where traditional optical methods are ineffective, such as in obscured or visually inaccessible environments. The potential of CSI-guided imaging to revolutionize applications in surveillance, autonomous navigation, and environmental sensing is immense, given its ability to operate under conditions that challenge conventional imaging systems.

However, a critical limitation of CSI-guided imaging is its inherent dependency on specific environmental characteristics, which poses a challenge to its adaptability across diverse and dynamically changing settings. The performance of CSI-guided imaging systems in accurately capturing dynamic objects and backgrounds is highly contingent on their ability to adapt to new environments—a capability that existing systems have not demonstrated.

In response to this gap, our study embarks on an experimental journey with the CSI-Imager, an extension of our previous CSI-Inpainter \cite{chen2023trans} system, aimed at exploring the feasibility of CSI-guided imaging across various environments through continuous model learning, specifically leveraging CL . Contrary to approaches reliant on transfer learning for adaptation, our work focuses on validating whether the CSI-Imager can maintain accurate imaging performance when transitioning between different environments, such as from office spaces to industrial settings, without the need for retraining or employing novel machine learning strategies.

This experimental study seeks to answer a pivotal question: Is adaptation for CSI-guided imaging across varying environments feasible through continuous model learning? By conducting a series of experiments—ranging from pretraining in an office environment to continuous updates in an industrial setting—we delve into the CSI-Imager's ability to adapt and perform in new contexts.

Our study's insights are notably applicable to the domain of vehicular technologies, particularly for scenarios within vehicles, where Wi-Fi signals are more consistently available and manageable. The internal environment of a vehicle presents a unique set of challenges for imaging systems due to its confined space and the presence of passengers and objects in constant motion. The CSI-Imager's ability to adapt and learn incrementally shows potential for enhancing in-vehicle safety systems, improving the precision of autonomous navigation systems, and supporting sophisticated traffic management strategies by providing dependable sensing capabilities within these confined settings.

This paper is organized as follows: Section II briefly reviews related work, setting the stage for our experimental focus. Section III details our methodology, emphasizing the design and data acquisition strategy of our CSI-guided imaging model. Section IV outlines the experimental setup and the CL strategy employed for model adaptation. Section V presents our preliminary experimental results, and discusses the CSI-Imager's performance in varied environments. In Section VI, we conclude by summarizing the study's contributions and outlining future research avenues to explore the adaptability of CSI-guided imaging systems further.

\section{Literature Review}
This section delves into the latest developments in wireless sensing technologies, with a spotlight on CSI-guided imaging. It further examines the pivotal role of CL in enabling these technologies to adapt to rapidly changing environments.

\subsection{Advancements in CSI-Guided Imaging}

Recent literature underscores the significant advancements in CSI-guided imaging, highlighting its potential across a broad spectrum of applications. Ma et al.'s foundational review of Wi-Fi sensing technologies elaborates on CSI's versatility in detection, recognition, and estimation tasks, suggesting the technology's applicability beyond human-centric sensing to include animals and inanimate objects \cite{ma2019wifi}. This broadening scope signals a transformative period in the evolution of wireless sensing.

Nalepa's exploration of advanced sensor technologies intersects directly with CSI-guided imaging, particularly emphasizing the role of multi- and hyperspectral imaging in augmenting wireless sensing capabilities \cite{s21186002}. Similarly, Garcia et al.'s investigation into optimized sensing matrices for compressive spectral imaging sensors presents critical insights into enhancing image reconstruction quality—a vital aspect of improving CSI-guided imaging systems \cite{garcia2020optimized}.

Innovative approaches to imaging, such as the CSI2Image method proposed by Kato et al., leverage GANs to convert CSI data into images, showcasing the efficacy of machine learning algorithms in refining the outputs of CSI-guided imaging systems \cite{9380376}. Furthermore, the pursuit of cost-effective and efficient sensing solutions by Wang et al. contributes to the ongoing development of CSI-guided imaging by emphasizing the importance of accessible technology \cite{wang2021low}.

\subsection{Role of CL in Overcoming CSI Domain Shift}

Adaptive methodologies have become essential in wireless sensing to address the challenges posed by domain shifts, particularly when leveraging CSI. The integration of CL offers a promising solution, enabling systems to dynamically adapt to new environments without extensive retraining. This subsection highlights significant contributions that showcase the potential of CL to mitigate CSI domain shifts in wireless contexts.

The survey by Chen et al. \cite{Chen2022Cross-Domain} underscores the difficulties faced by CSI-based sensing systems when transitioning between domains, advocating for algorithms that enhance sensing accuracy amidst such shifts. This research emphasizes the critical role of CL in facilitating system adaptability across diverse environments.

Berlo et al. \cite{Berlo2023Mini-Batch} introduce an innovative technique of mini-batch alignment for domain-independent feature extraction from Wi-Fi CSI data. By guiding the model's training process, this method effectively addresses domain shifts, underscoring the versatility of CL in wireless sensing applications.

Zhu et al. \cite{Zhu2022Wireless} contribute to the discourse with their CNN-based wireless channel recognition algorithm, which leverages multi-domain feature extraction to maintain performance under varying conditions. Their work exemplifies how learning-based methods can navigate the intricacies of domain shifts, further validating the efficacy of CL strategies.

Furthermore, Du et al. \cite{Du2024Multidomain} highlight the need for advanced data solutions in device-free Wi-Fi sensing that exploit CSI's multidimensional nature. Their approach to generating multidimensional tensors for finer-grained contextual information extraction illustrates a forward-thinking application of CL to enhance indoor scenario characterization.

Complementary insights are provided by Cho et al. \cite{Cho2023Complementary}, who explore unsupervised continual domain shift learning with their Complementary Domain Adaptation and Generalization (CoDAG) framework, aiming to achieve system adaptability and memory retention across domains. Similarly, Simon et al. \cite{9879030}, Houyon et al. \cite{10208897}, and Van Berlo et al. \cite{Berlo2023Use} discuss methodologies that mitigate catastrophic forgetting and enhance cross-domain performance, reinforcing the significance of CL in the context of CSI domain shifts.

Liu and Ding \cite{Liu2022Training} offer a perspective on training enhancement for Deep Learning (DL) models in CSI feedback mechanisms, emphasizing the exploitation of CSI features to augment dataset capabilities—a strategy aligned with the principles of CL .

Collectively, these studies underscore the evolving landscape of wireless sensing, where CL emerges as a crucial mechanism to address CSI domain shifts. By adopting strategies like mini-batch alignment and multi-domain feature extraction, the field is advancing towards more resilient and adaptable wireless sensing systems, capable of thriving in the dynamically changing real-world environments.

\section{Methodology of CSI-Guided Imaging}

Leveraging the potential of wireless sensing technologies, CSI-Imager transforms CSI data into visual images, harnessing a combination of wireless channel characteristics and DL techniques \cite{chen2023trans}. This section outlines the methodology underpinning CSI-Imager, focusing on its system components, data preprocessing techniques, and the DL architecture designed for CSI-guided imaging.

\subsection{System Model}

\begin{figure*}
    \centering
    \includegraphics[scale=0.4]{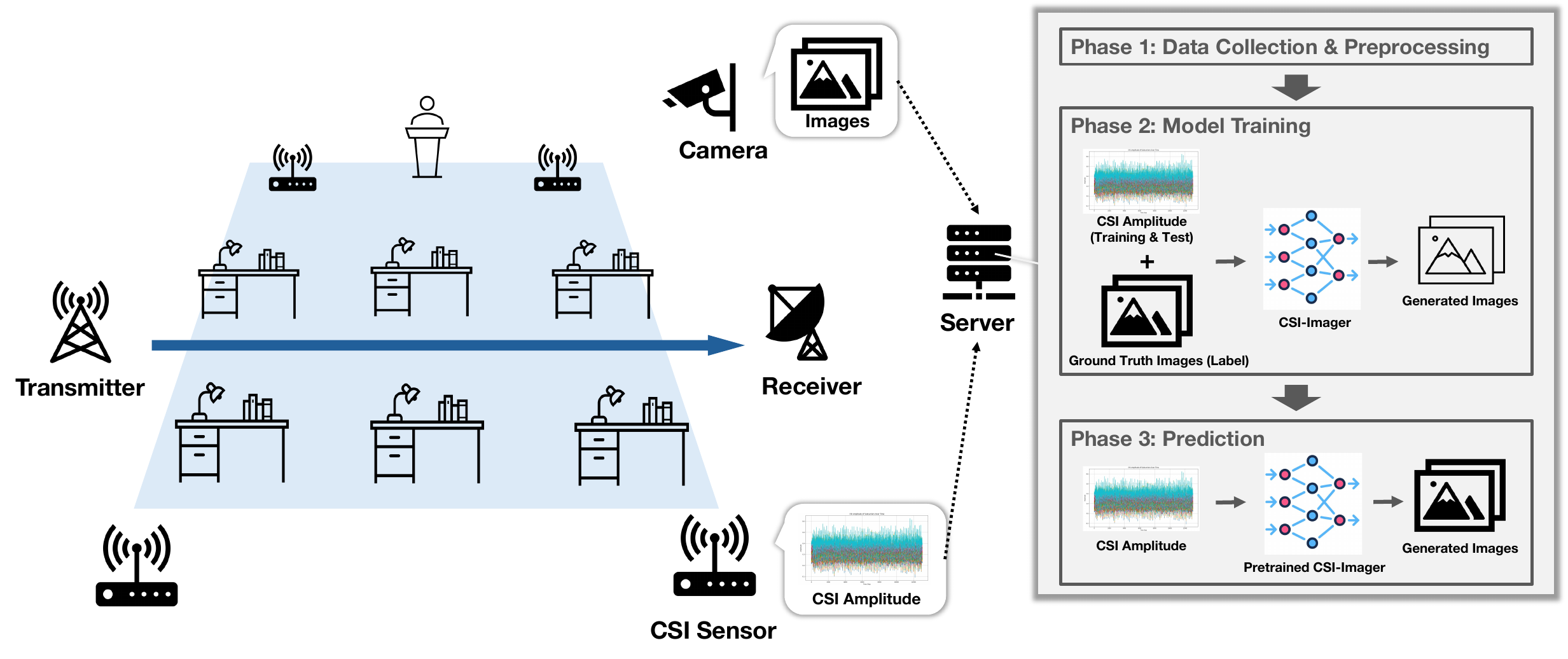}
    \caption{System model of CSI-Imager.}
    \label{fig:system_model}
\end{figure*}

The CSI-Imager system encompasses cameras, CSI sensors, a data preprocessing module, and a bespoke deep neural network architecture (see \cref{fig:system_model}). Cameras and CSI sensors collaborate to capture visual and CSI from the environment. While cameras record RGB images, CSI sensors gather data reflecting the wireless channel's behavior, influenced by environmental obstacles. This data synergy aids in reconstructing comprehensive images of the environment, addressing areas occluded or otherwise missing in the visual data. Data preprocessing is crucial for synchronizing and refining both image and CSI data before model training and prediction. The DL component, central to the CSI-Imager, employs a sophisticated architecture comprising an encoder and a decoder to transform CSI into images.

\subsection{Data Collection and Preprocessing}

\subsubsection{Data Collection}

CSI is collected using Long Training Symbols from packet preambles in a MIMO-OFDM Wi-Fi system. This process yields a complex-valued CSI matrix, encapsulating the amplitude attenuation and phase shift across spatial, frequency, and temporal dimensions. For imaging purposes, the focus is on the amplitude component of CSI, which provides critical signal strength information across spatial and temporal modes, while the phase component is omitted to simplify the analysis.

Synchronization between image frames and CSI sequences is ensured through a unified clock across devices within a local network, facilitated by the Network Time Protocol, ensuring temporal alignment at a collection rate of $10$ fps.

\subsubsection{Data Preprocessing}

Following collection, image and CSI data undergo rigorous preprocessing to remove noise and align data points temporally. Images are resized to standard dimensions, and a low-pass filter is applied to CSI data to enhance quality. Isochronization of data ensures each CSI matrix precisely corresponds to its related image frame, leveraging a bisection search method to align time stamps accurately.

\subsection{CSI-Imager Model Architecture}

\begin{figure*}
    \centering
    \includegraphics[scale=0.4]{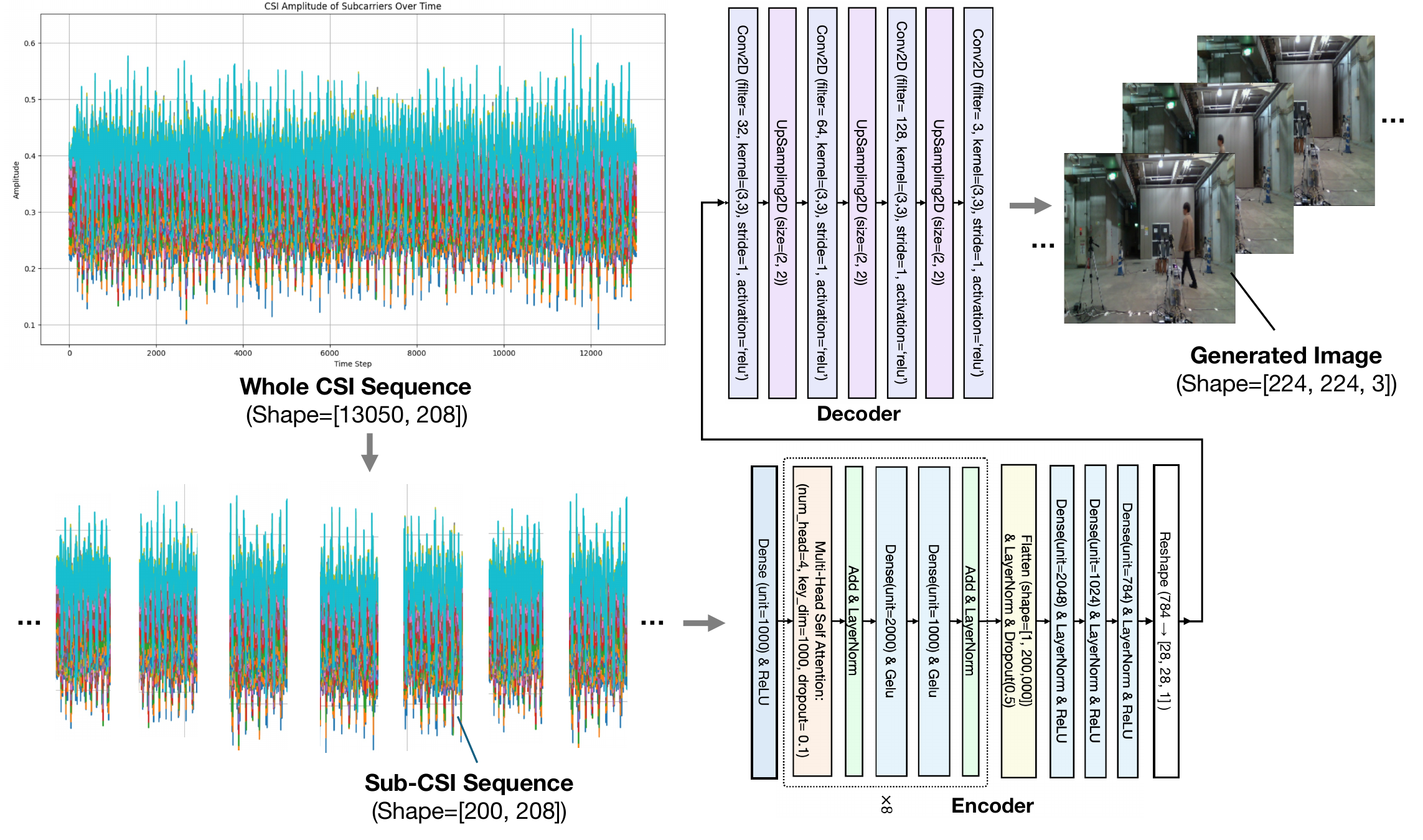}
    \caption{The model architecture of CSI-Imager.}
    \label{fig:model}
\end{figure*}

The CSI-Imager's architecture, illustrated in \cref{fig:model}, is partitioned into an encoder and a decoder. The encoder adeptly processes the CSI data, extracting essential features using a Transformer mechanism known for its efficiency in handling sequential data. The decoder then reconstructs visual images from these encoded features through convolutional layers and upsampling.

\subsection{Training and Prediction Procedures}

Training CSI-Imager involves pairing image sequences with corresponding CSI matrices, where the DL model learns to correlate CSI-derived features with visual information. This process aims to generate accurate visual representations from CSI data, enhancing the capability of wireless sensing in imaging applications. The model's effectiveness is subsequently evaluated on a validation set to ensure reliable performance across different environmental settings.

\section{CL Strategy for Environmental Adaptation}

This section outlines the strategy employed to facilitate CSI-Imager's adaptation across dynamically changing environments through CL. This approach is designed to ensure the model's ongoing evolution and knowledge retention, critical for its application in real-time CSI-guided imaging.

\subsection{Adapting Through CL }

Central to CSI-Imager's adaptability is CL , enabling it to seamlessly adapt to dynamically changing environments. This strategy empowers the system to absorb and integrate new environmental data in real-time, enhancing its performance across varied conditions. By employing online learning algorithms, CSI-Imager dynamically updates its parameters with each new data batch, facilitating continuous evolution without losing previously acquired knowledge. This approach ensures sustained imaging performance and adaptability, marking CSI-Imager as a highly effective tool in navigating the complexities of different settings.

\subsection{Adaptation Mechanisms}

To support the CL approach, several mechanisms are integrated into the CSI-Imager framework, promoting its adaptability and sustained performance:

\begin{enumerate}
\item \textit{Real-time Data Integration}: A continuous influx of environmental CSI data is processed to ensure the model remains up-to-date with current conditions.
\item \textit{Feedback Loop for CL }: Model parameters are dynamically updated based on real-time data analysis, ensuring the model remains aligned with the latest environmental characteristics.
\item \textit{Performance Monitoring and Adjustment}: The imaging quality is continuously monitored against predefined metrics, with model parameters fine-tuned as needed to maintain or enhance performance.
\item \textit{Activation of CL Modules}: Based on the detected environmental characteristics, specific CL modules are activated to address the unique challenges of each environment.
\end{enumerate}

This strategy ensures that CSI-Imager not only adapts swiftly to new environments but also retains and continuously builds upon its learned knowledge, showcasing enhanced flexibility and accuracy in CSI-guided imaging across varying conditions.

\subsection{CL Algorithm for Environmental Adaptation}

CSI-Imager adapts to dynamically changing environments through a structured CL process, leveraging real-time CSI data to maintain high imaging performance. This adaptation is guided by evaluating image quality through Mean Structural Similarity Index Measure (SSIM) and Mean Peak Signal-to-Noise Ratio (PSNR), offering a quantitative assessment of imaging fidelity. Below, we detail the definitions and algorithm driving this adaptation process. 

Definitions for variables:
\begin{itemize}
\item $D^\mathrm{CSI}_t$: CSI data samples collected at time slot $t$.
\item $D^\mathrm{Img}_t$: Corresponding image data samples collected at time slot $t$.
\item $M_t$: The model state at time slot $t$.
\item $I_t$: Predicted image by CSI-Imager using $D^\mathrm{CSI}_t$.
\item $S(I_t, D^\mathrm{Img}_t)$: Mean image quality score of model $M$ for dataset $D$, utilizing metrics such as PSNR and SSIM.
\item $S_\mathrm{th}$: Threshold for required image quality score.
\end{itemize}

\begin{algorithm}
\caption{CL Strategy for CSI-Imager}
\begin{algorithmic}[1]

\State \textbf{Training Process:}
\State Initialize $M_0$ with a pre-trained CSI-Imager model.
\For{each time slot $t$}
\State $I_t \gets M_{t-1}(D^\mathrm{CSI}_t)$.
\If{$S(I_t, D^\mathrm{Img}_t) < S_\mathrm{th}$}
\State Update $M_{t-1}$ with $(D^\mathrm{CSI}_t, D^\mathrm{Img}_t)$ in a supervised learning manner to obtain $M_t$.
\Else
\State $M_t \gets M_{t-1}$.
\EndIf
\EndFor
\\

\State \textbf{Inference Process:}
\For{each time slot $t$}
    \If{inference request comes}
        \State Copy the latest model, $M_t$, from the training process.
        \State \textbf{Output:} $M_t(D^\mathrm{CSI}_t)$.
    \EndIf
\EndFor
\end{algorithmic}
\end{algorithm}

The CL strategy consists of two primary processes: training and inference, which highlights the iterative process through which CSI-Imager systematically updates its model to accommodate new environmental conditions. By evaluating and adapting the model in alignment with the Mean SSIM and Mean PSNR scores, the system ensures effective adaptation and high fidelity in imaging across diverse environmental contexts, thus maintaining optimal performance.

Building upon the described strategy for environmental adaptation, our experimental evaluation specifically assumes a scenario where $S_\mathrm{th}$ is set at a significantly high value (Mean SSIM: $0.9$, Mean PSNR: $28~dB$). This assumption is critical as it ensures the model is in a constant state of update, adapting continuously to the incoming stream of environmental CSI data.

\section{Domain Adaptation and Imaging Performance}

\subsection{Experimental Setup}

\begin{table}[t]
\caption{Experimental equipment}
\label{tab:exp_equip}
\centering
\begin{tabular}{cc} 
\toprule
Receiver            & NETGEAR Nighthawk X10 \\
Transmitter         & NETGEAR Nighthawk X10 \\
Wireless LAN standard & IEEE 802.11ac                \\
Channel             & 36                    \\
Bandwidth           & 80\,MHz               \\ 
\midrule
CSI sensor          & Raspberry Pi 4 model B       \\
CSI sensor firmware & Nexmon CSI            \\
CSI measurement rate & 500\,Hz \\ 
\midrule
Camera 1,2          & RealSense L515        \\
Camera 3            & RealSense D435       \\ 
\bottomrule
\end{tabular}
\end{table}

\begin{figure*}
    \begin{tabular}{cc}
    
      \begin{minipage}{\linewidth}
        \centering
        \includegraphics[scale=0.25]{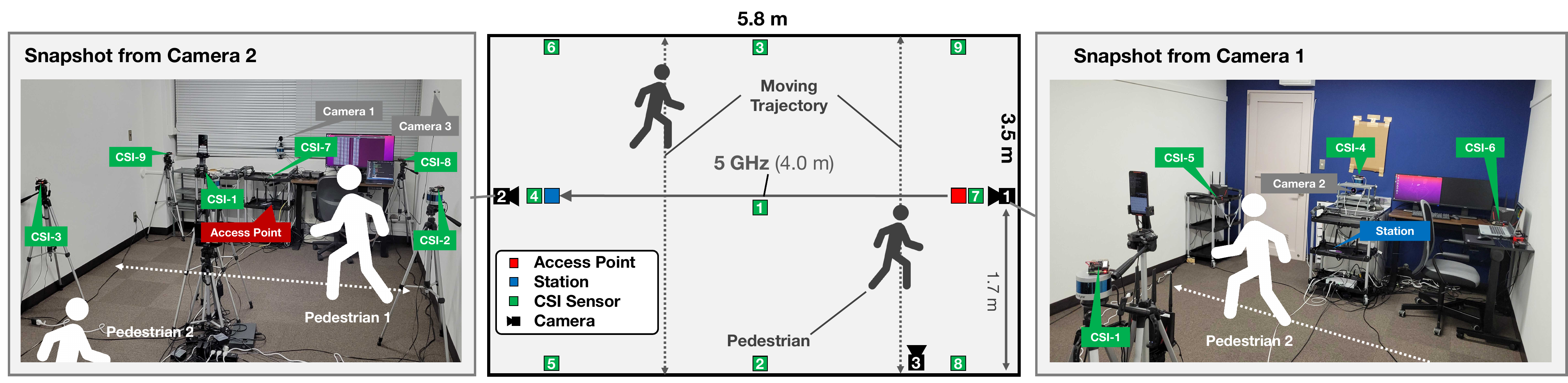}
        \subcaption{ Configuration for the office experiment.}
        \label{fig:envir01}
      \end{minipage}\\

      \begin{minipage}{\linewidth}
        \centering
        \includegraphics[scale=0.25]{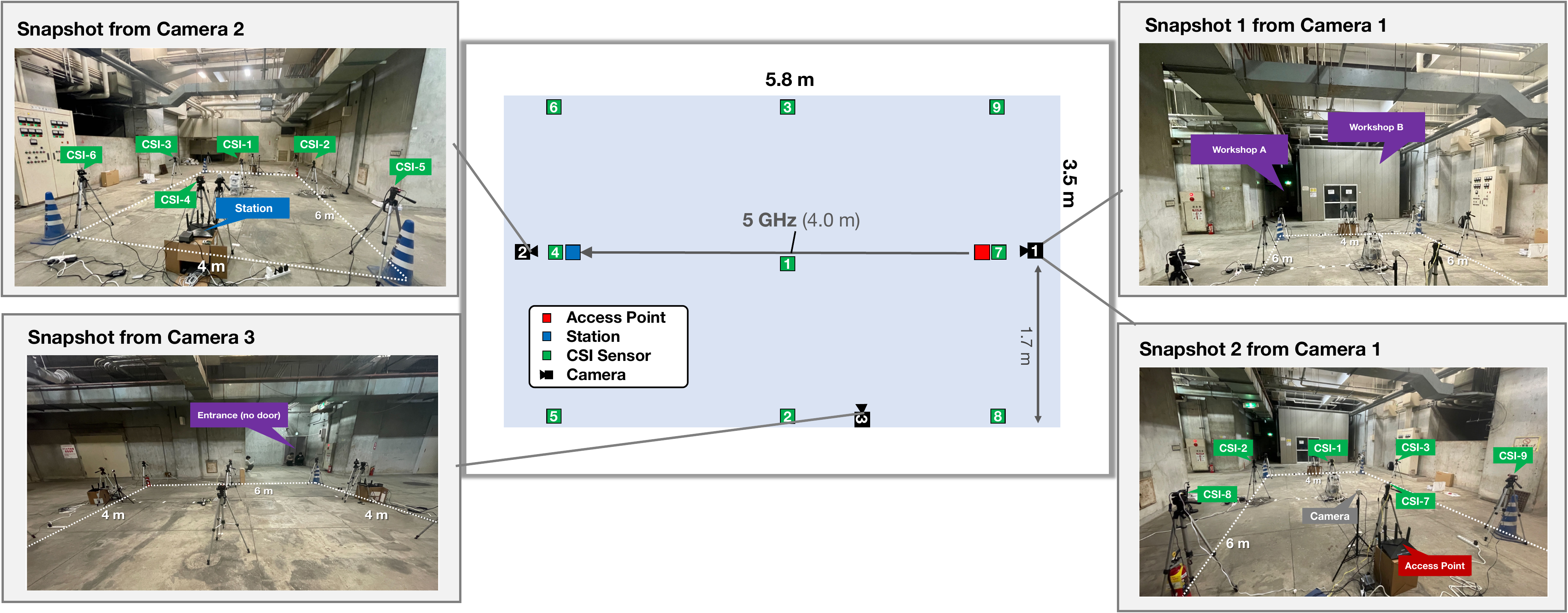}
        \subcaption{ Configuration for the industrial experiment.}
        \label{fig:envir02}
      \end{minipage}
      
    \end{tabular}
    \caption{ The experimental setup.}
    \label{fig:offi_envir}
\end{figure*}

\subsubsection{Settings for Office Experiment}

The first experiment was conducted in an office environment, where pedestrians periodically obstructed the line-of-sight path of a wireless LAN connection. Equipment used and the experimental setup are detailed in \cref{tab:exp_equip}, and the configuration is depicted in \cref{fig:envir01}.

To generate traffic, wireless LAN devices were placed at both ends of the room, using iperf as the traffic generator. Nine CSI sensors placed in the environment captured the wireless LAN signal, collecting CSI data. The movements of pedestrians caused variations in CSI values, which were captured by three RGB cameras.

The experiment lasted 30 minutes, resulting in temporally continuous sequences of RGB images, CSI matrices from different sensors. Each sequence contained 18,000 entries, providing a comprehensive dataset for analysis.

\subsubsection{Settings for Industrial Experiment}

To rigorously assess the CSI-Imager's adaptability and robustness, a series of experiments were conducted in a factory workshop, a setting marked by its complexity and challenging conditions (as depicted in \cref{fig:envir02}). This environment introduces potential noise interference from adjacent machinery and unpredicted disturbances from non-participant pedestrian traffic. Furthermore, the spacious and semi-open nature of the workshop potentially compromises the integrity of reflected Wi-Fi signals, posing a challenge to the accuracy of CSI data acquisition due to signal attenuation in open-air conditions.

In an effort to minimize variables and maintain consistency with prior experiments, all equipment used in the office setting was relocated to a similarly sized rectangular area within the factory space, measuring $5.8~m \times 3.5~m$.

The experimental protocol was designed to mimic a range of real-life situations through a series of progressively complex scenarios, from individual to group activities. To ensure data variability and enhance result reliability, five participants were dressed distinctly and instructed to alternate their walking direction between scenarios. The scenarios unfolded as follows:
\begin{itemize}
\item \textbf{Scenario 1 (S1):} Featured a single participant (P1) moving in a clockwise direction within the designated area for a duration of 10 minutes.
\item \textbf{Scenario 2 (S2):} Directly followed S1 with participant (P2) traversing the area in a counterclockwise direction for 10 minutes, introducing an immediate shift in pedestrian dynamics.
\item \textbf{Scenario 3 (S3):} Continued the sequence with a third individual (P3) adopting a clockwise movement pattern for another 10 minutes, seamlessly extending the data collection phase.
\item \textbf{Scenario 4 (S4):} A composite scenario that saw the combination of P1, P2, and P3 navigating the area in unison, clockwise for 10 minutes, to simulate group dynamics.
\item \textbf{Scenario 5 (S5):} Introduced a new assembly of individuals (P1, P4, and P5), exploring the space counterclockwise for an additional 10 minutes, further complicating the environmental variables.
\item \textbf{Scenario 6 (S6):} Culminated with an inclusive session involving all five participants (P1 through P5) independently maneuvering within the area in varying directions for 20 minutes, representing the peak of environmental complexity.
\end{itemize}

The CSI and image data collection spanned approximately two hours, encompassing S1 through S6 in a seamless and uninterrupted manner. This methodical and continuous data acquisition strategy was deliberately designed to rigorously assess the CSI-Imager's performance across a spectrum of environmental and pedestrian dynamics, ensuring a thorough evaluation of its adaptability and imaging accuracy in real-world settings.

\subsection{Pre-training on Office Environment Dataset}

\begin{figure}
\centering
\includegraphics[scale=0.4]{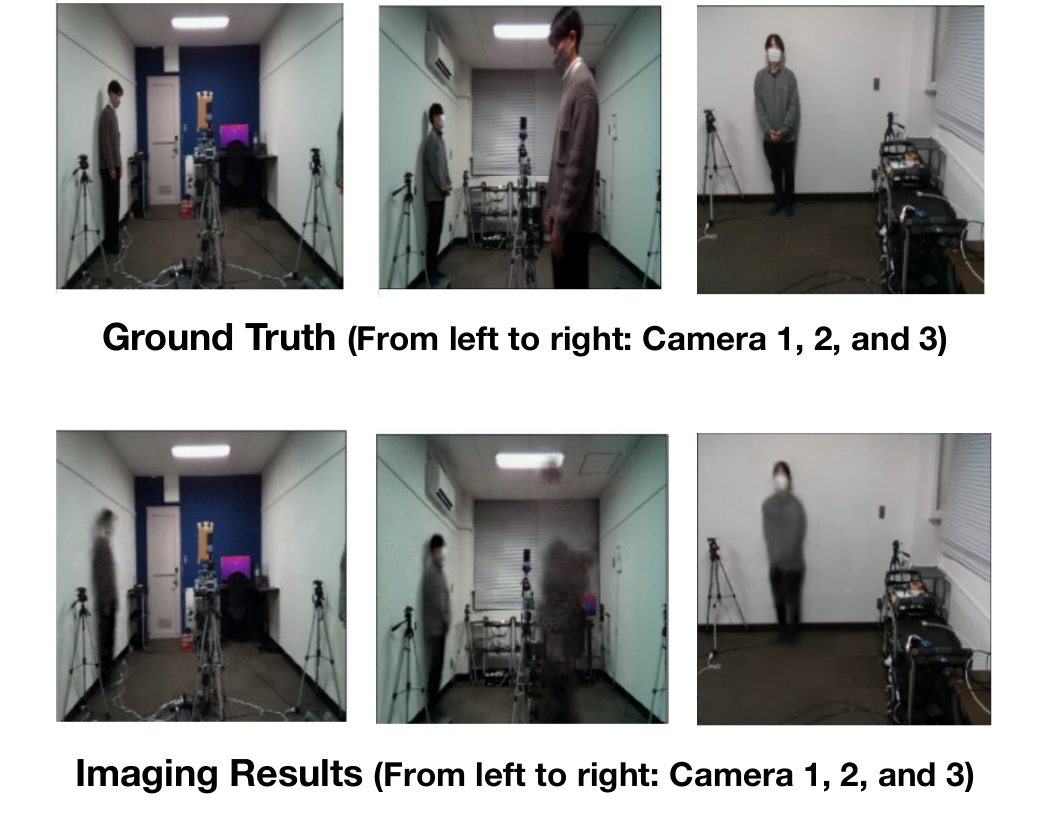}
\caption{Sample imaging results in office environments for Camera 1, Camera 2, and Camera 3, illustrating the CSI-Imager's initial training performance.}
\label{figs:office_result}
\end{figure}

The adaptability of CSI-Imager is rigorously tested as we transition from the office setting to the industrial workshop. Initially, the CSI-Imager with four CSI input channels is pretrained on datasets collected from each camera in the office environment respectively, undergoing a comprehensive training process over 50 epochs. This pretraining phase is critical for the model to learn and internalize the specific patterns and nuances of the office setting, thereby establishing a robust baseline for imaging performance. Sample imaging results from this phase, demonstrating the model's proficiency across Cameras 1, 2, and 3, are depicted in \cref{figs:office_result}.


\subsection{Qualitative Evaluation of Model Adaptation via Continual Learning}

We experimentally evaluate the adaptation of models to scenarios different from the pre-trained environment using CSI-Imager. For each scenario, the model is fine-tuned for a fixed number of epochs on the dataset specific to that scenario. Subsequently, images are generated for validation samples acquired from the same scenario. The quality of these generated images is qualitatively assessed to demonstrate the model's adaptability to new environments.

\begin{figure}
   \centering
    \includegraphics[scale=0.3]{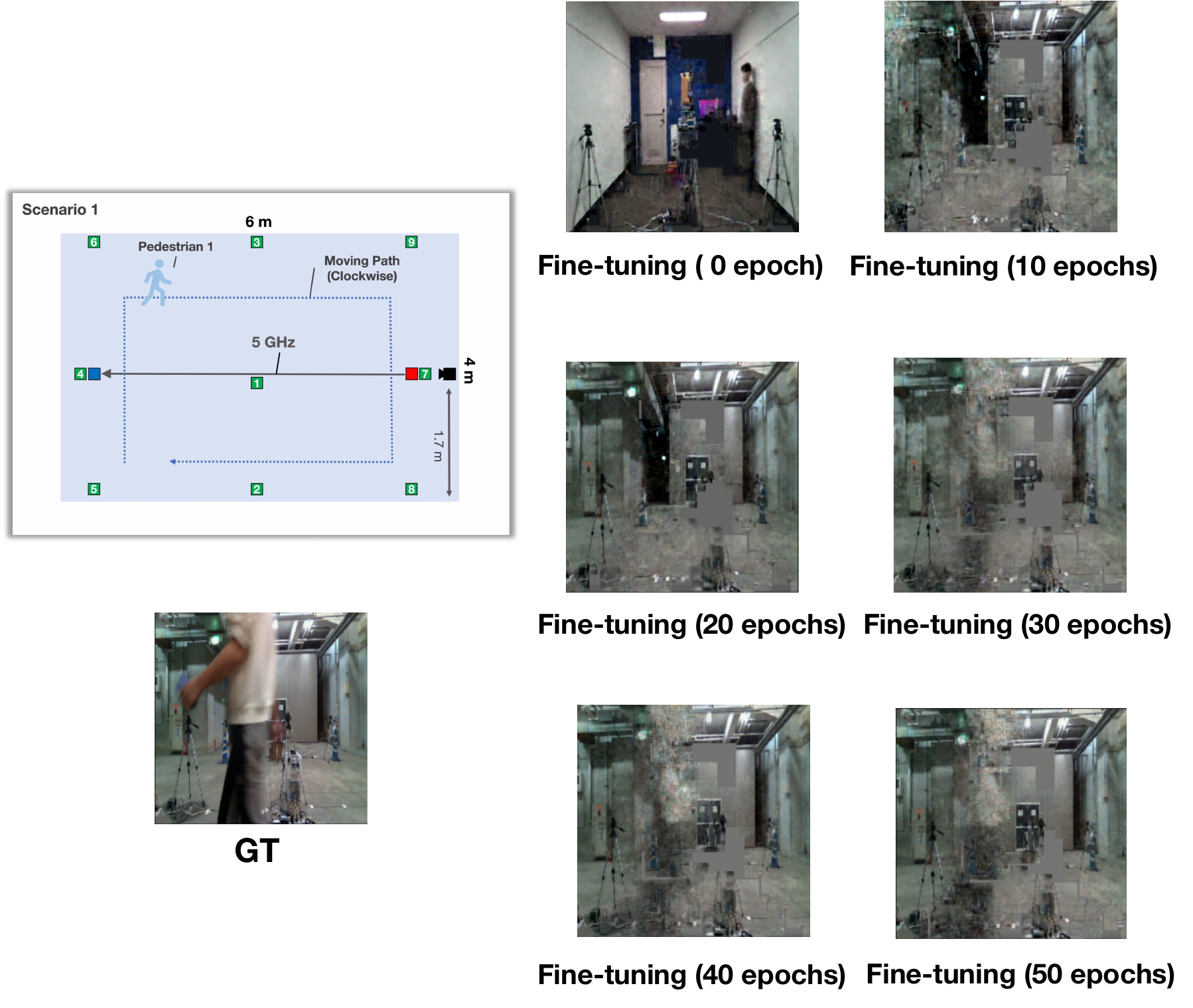}
    \caption{Adaptation results of S1.}
    \label{fig:factory_result_01}
\end{figure}

\cref{fig:factory_result_01} illustrates the adaptation process from an office environment to an industrial environment, specifically to scenario S1.
As depicted in \cref{fig:factory_result_01}, initial fine-tuning was essential, with the model undergoing at least ten epochs to adapt to the distinct characteristics of the new environment. Notably, after a 30-epoch tuning period, the dynamic representation of P1 was significantly improved, indicating that further updates beyond 50 epochs yielded diminishing returns. Consequently, for subsequent scenarios, the model updating ceased after 30 epochs to optimize efficiency.

\begin{figure}
   \centering
    \includegraphics[scale=0.3]{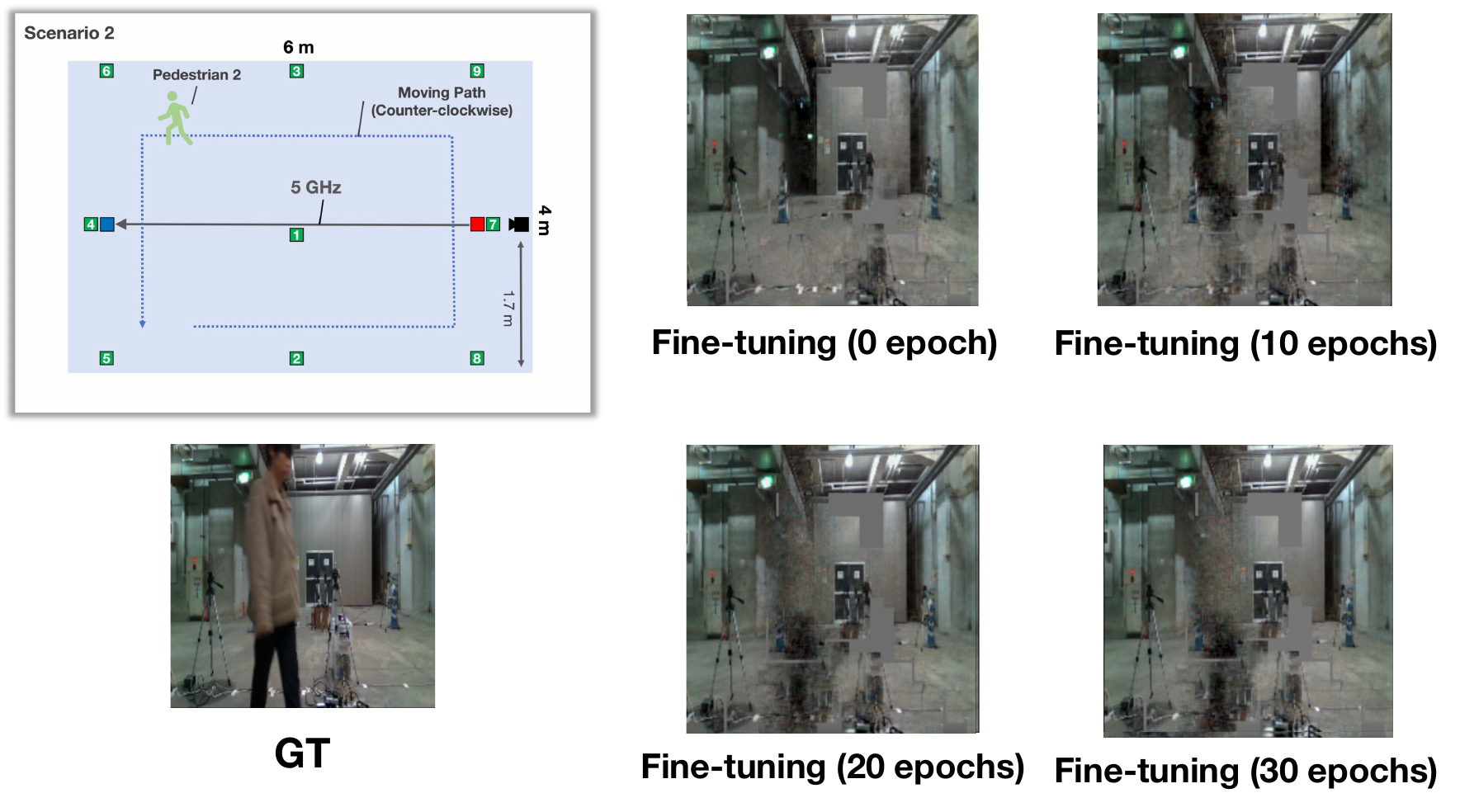}
    \caption{Adaptation results of S2.}
    \label{fig:factory_result_02}
\end{figure}

\begin{figure}
   \centering
    \includegraphics[scale=0.3]{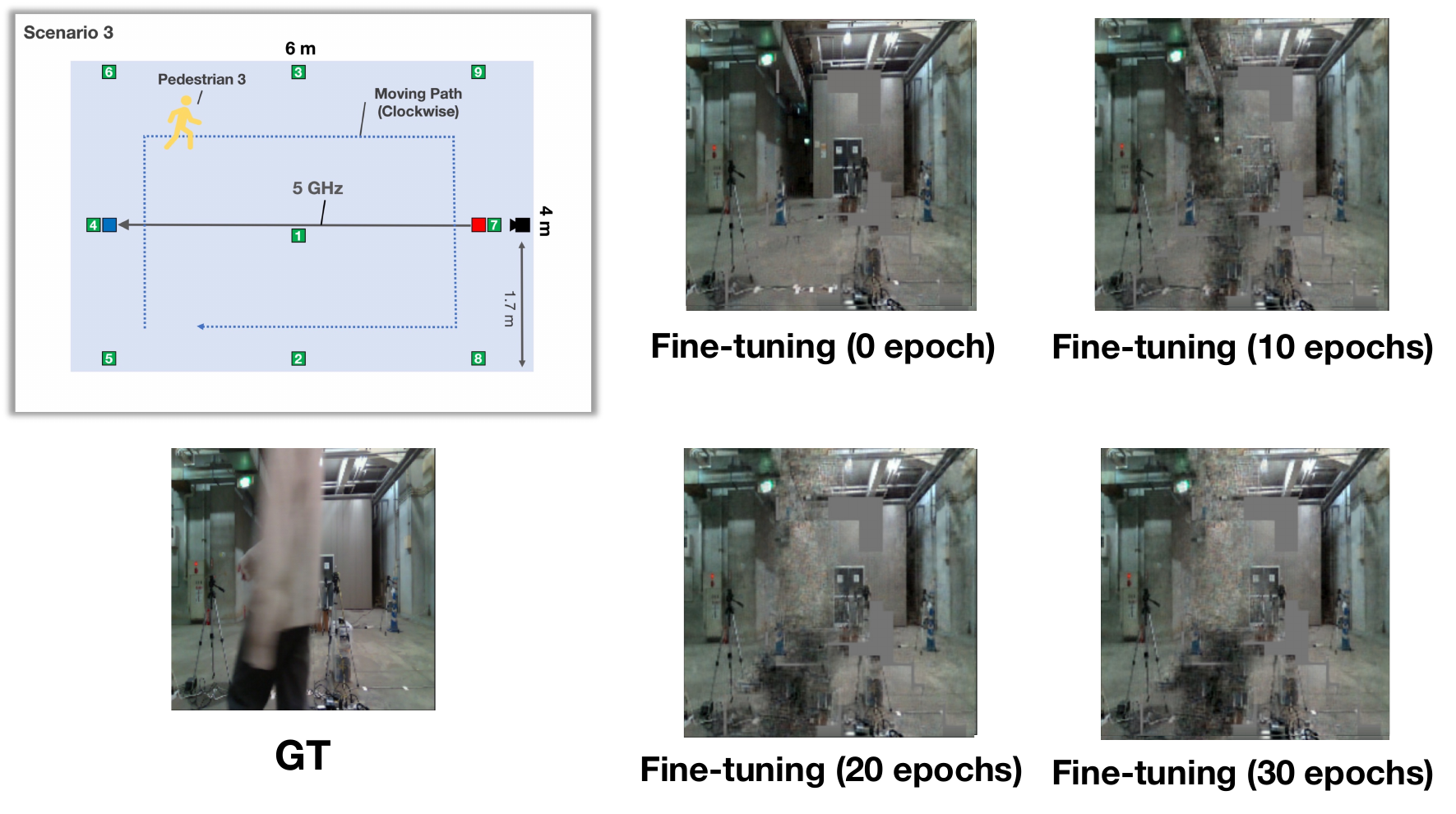}
    \caption{Adaptation results of S3.}
    \label{fig:factory_result_03}
\end{figure}

Further analysis in single-pedestrian scenarios (S2 and S3) revealed a more rapid adaptation process, underscoring the model's capability for swift learning and adjustment to novel environmental conditions, as illustrated in \cref{fig:factory_result_02} and \cref{fig:factory_result_03}.

\begin{figure}
   \centering
    \includegraphics[scale=0.3]{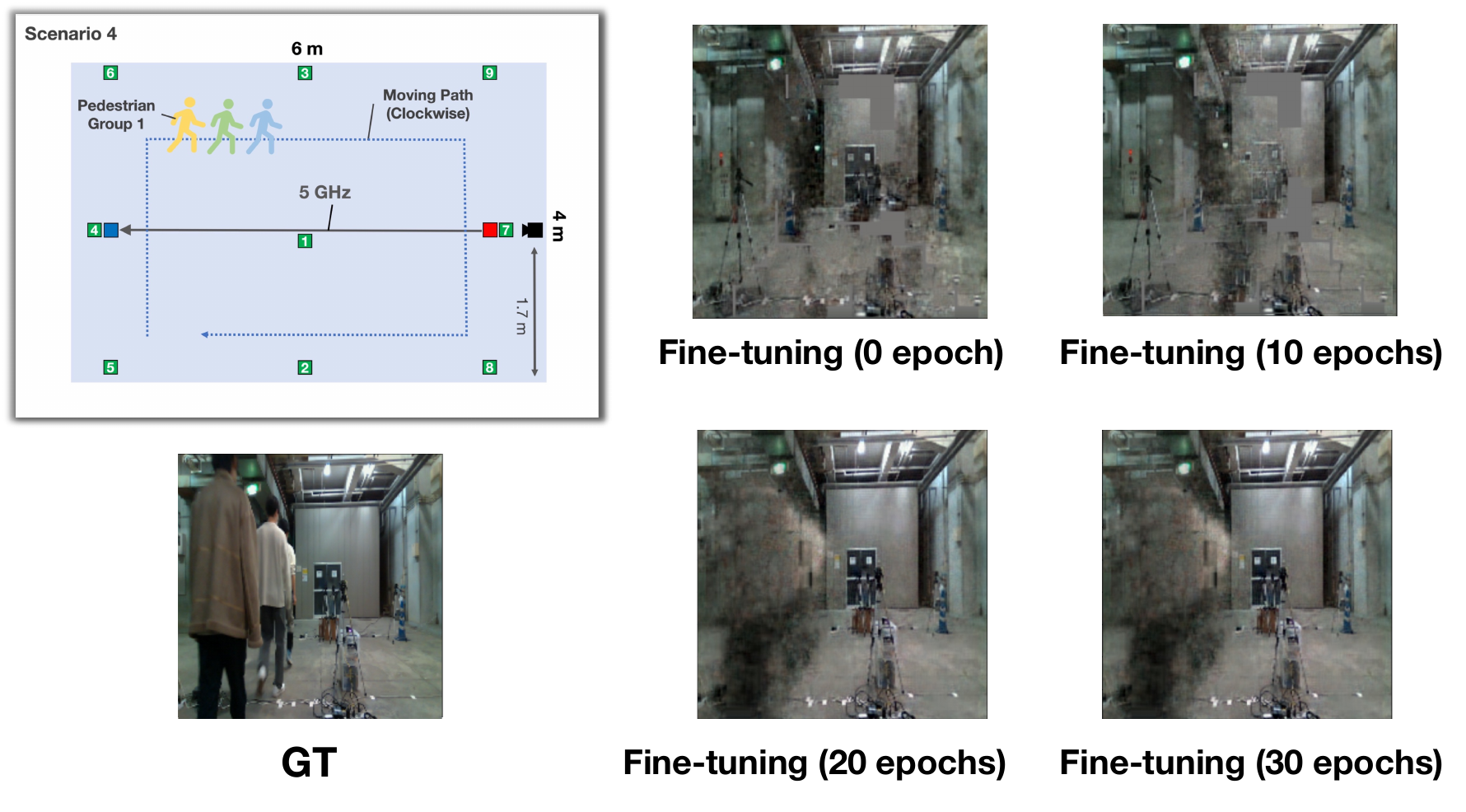}
    \caption{Adaptation results of S4.}
    \label{fig:factory_result_04}
\end{figure}

\begin{figure}
   \centering
    \includegraphics[scale=0.3]{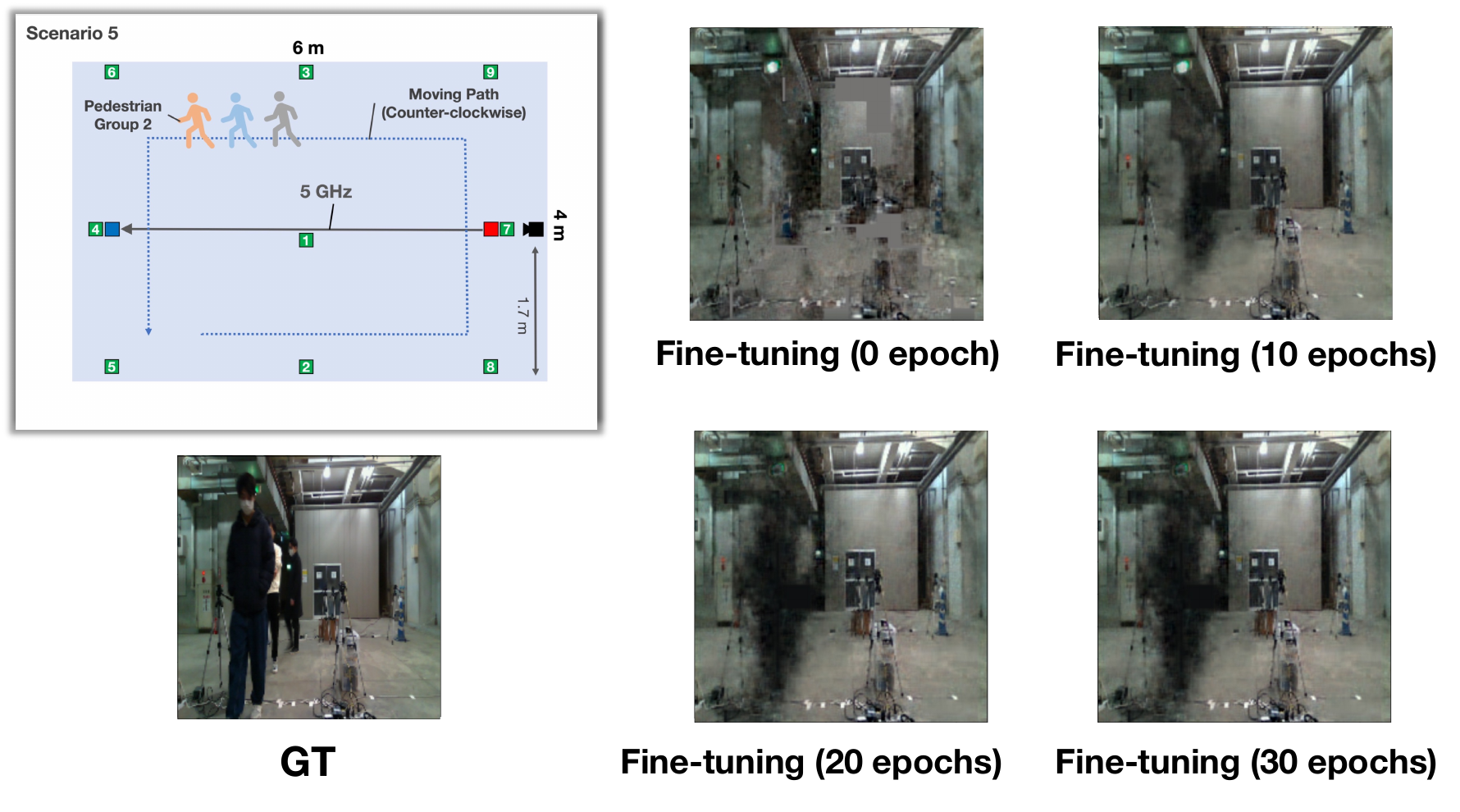}
    \caption{Adaptation results of S5.}
    \label{fig:factory_result_05}
\end{figure}

\begin{figure}
   \centering
    \includegraphics[scale=0.3]{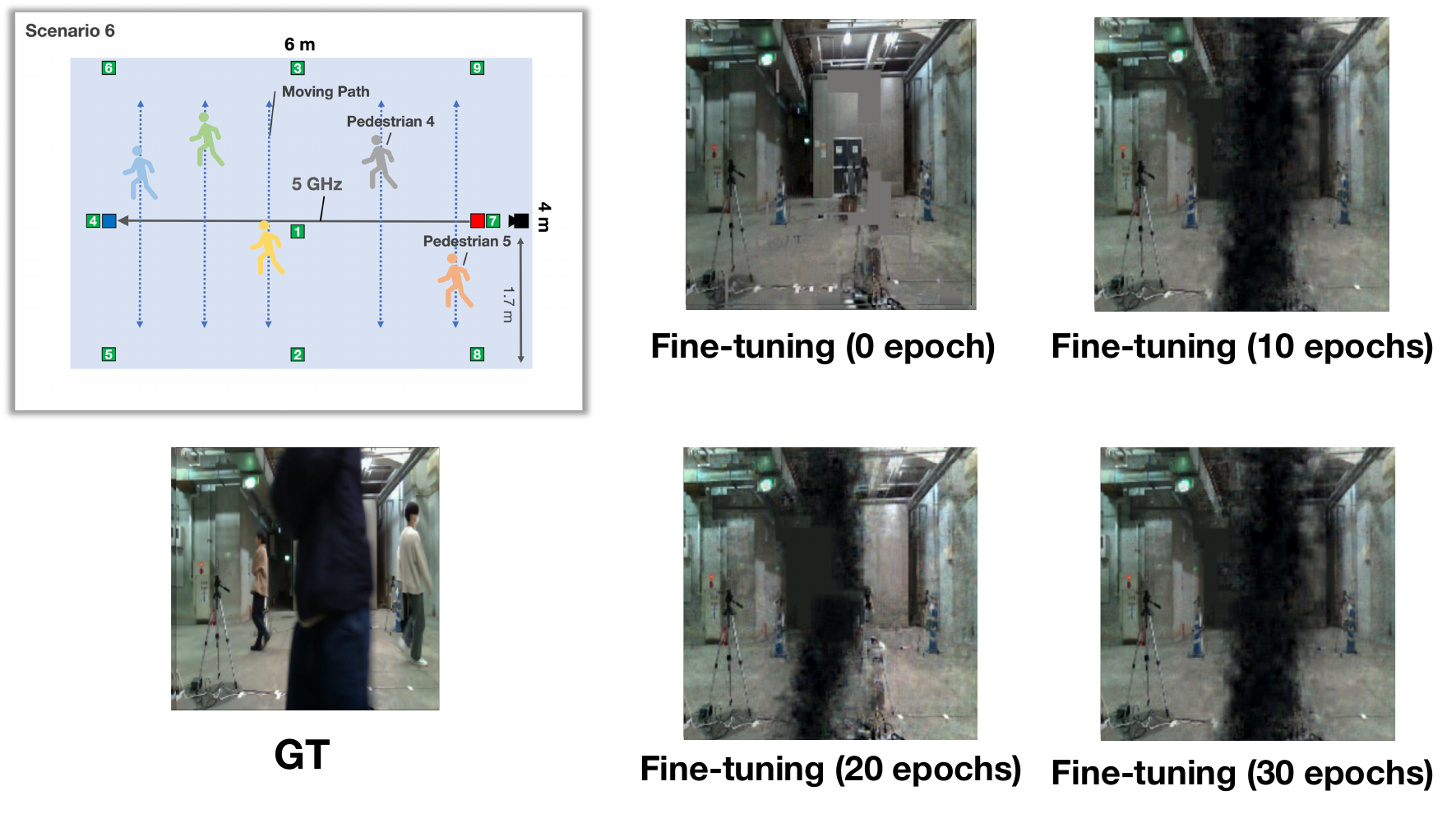}
    \caption{Adaptation results of S6.}
    \label{fig:factory_result_06}
\end{figure}

The challenge escalated with multi-pedestrian scenarios (S4-S6), introducing a higher complexity to the imaging task. Despite these added complexities, the CSI-Imager adeptly reconstructed the color and location of all subjects, showcasing its effectiveness across scenarios as seen in \cref{fig:factory_result_04}, \cref{fig:factory_result_05}, and \cref{fig:factory_result_06}. Although S6 presented additional challenges—predominantly the model's preference for imaging subjects closer to the camera—the successful outcomes across these scenarios reinforce the CSI-Imager's applicability in a variety of industrial and dynamic environments where conventional imaging methods are less effective.

Transitioning from an office to an industrial setting, CSI-Imager not only demonstrated its adaptability through CL but also affirmed its superior capability in RF-based imaging. Concentrating on its performance through diverse and demanding scenarios, CSI-Imager establishes a new standard for accuracy in visual representation within dynamic environments, highlighting its potential for extensive application in real-world settings.

\section{Conclusion}
In this paper, we present preliminary experimental results that demonstrate the potential for adapting the CSI-Imager, a system designed for imaging environments from Wi-Fi CSI, to changes in environment and scenario through continual learning. 
Through a series of experiments, we transitioned CSI-Imager from a familiar office environment to the uncharted territories of an industrial setting, closely observing its performance and adaptability through continuous model updates.


Future research directions include conducting experiments in vehicular environments, particularly demonstrating the feasibility of imaging within vehicles, and exploring CSI-assisted Non-Line-of-Sight (NLOS) imaging to visualize areas that are blind spots for cameras using supplementary information from CSI. Additionally, the development of new algorithms for more efficient transfer and CL processes could further optimize the adaptability and performance of CSI-guided imaging systems in dynamically changing environments.

\section*{Acknowledgment}
This research was funded by the JSPS KAKENHI Grant Number JP22H03575.

\bibliographystyle{IEEEtran.bst}
\bibliography{main_text.bib}

\end{document}